\documentclass[a4paper,11pt,]{article}
\pdfoutput=1 
\usepackage{jcappub} 
\usepackage{printlen}
\uselengthunit{in}
\usepackage{dsfont}
\usepackage{amsmath}
\DeclareMathOperator*{\argmax}{arg\,max}
\DeclareMathOperator{\sinc}{sinc}
\usepackage{float}
\usepackage{bm}
\def\apss{\ref@jnl{Ap\&SS}}             

\usepackage[T1]{fontenc} 
\usepackage{mathtools}
\usepackage{lipsum}
\usepackage{graphicx}
\usepackage[usenames, dvipsnames]{xcolor}

\usepackage{color}
\definecolor{darkgreen}{rgb}{0,0.7,0}

\newcommand{\neww}[1]{{#1}}
\definecolor{c1}{RGB}{249,65,68} 
\definecolor{c4}{RGB}{255,111,114} 
\definecolor{c2}{RGB}{0,168,50} 
\definecolor{c3}{RGB}{39,125,161} 
\definecolor{c5}{RGB}{157,111,255} 
\definecolor{c6}{RGB}{251,105,255} 

\newcommand{\vOm} { {\bm{\Omega}} }
\newcommand{\vom} { {\bm{\omega}} }

\newcommand{\vw} { {\bm{w}} }
\newcommand{\vY} { {\bm{Y}} }
\newcommand{\vN} { {\bm{N}} }

\newcommand{\vx} {\bm{x}}
\newcommand{\vq} { {\bm{q}} }

\newcommand{\rd}[2]{\frac{d #1}{d#2}}

\renewcommand{\vN}{\mathbf{N}}

\newcommand{\vtheta} {{\bm{\theta}}}

\newcommand{\aap}{Astron.\ Astrophys.\ }

\newcommand{\apjl}{Astrophys.\ J.\ Lett.\  }
\newcommand{\apjs}{Astrophys.\ J.\ Suppl.\  }

\DeclareRobustCommand{\Sec}[1]{Sec.~\ref{sec:#1}}

\DeclareRobustCommand{\App}[1]{App.~\ref{app:#1}}

\DeclareRobustCommand{\Tab}[1]{Table~\ref{tab:#1}}

\DeclareRobustCommand{\Fig}[1]{Fig.~\ref{fig:#1}}
\DeclareRobustCommand{\Figs}[2]{Figs.~\ref{fig:#1} and \ref{fig:#2}}
\DeclareRobustCommand{\Eq}[1]{Eq.~(\ref{eq:#1})}
\DeclareRobustCommand{\Eqs}[2]{Eqs.~(\ref{eq:#1}) and (\ref{eq:#2})}
\DeclareRobustCommand{\Reff}[1]{Ref.~\cite{#1}}
\DeclareRobustCommand{\Refs}[1]{Refs.~\cite{#1}}

\title{\boldmath Aemulus \nu}

\author[a,b]{Delon Shen}
\emailAdd{delon@stanford.edu}
\author[c,d]{Nickolas Kokron}
\emailAdd{kokron@astro.princeton.edu}
\author[e]{Joseph DeRose}
\emailAdd{jderose@lbl.gov}
\author[f]{Jeremy Tinker}
\emailAdd{tinker@nyu.edu}
\author[a,b]{Risa H. Wechsler}
\emailAdd{rwechsler@stanford.edu}
\author[g]{Arka Banerjee}
\emailAdd{arka@iiserpune.ac.in}
\author[]{and the Aemulus Collaboration}

\affiliation[a]{Department of Physics, Stanford University, 382 Via Pueblo Mall, Stanford, CA 94305, USA}
\affiliation[b]{Kavli Institute for Particle Astrophysics and Cosmology, SLAC National Accelerator Laboratory, 2575 Sand Hill Road, Menlo Park, CA 94025, USA}
\affiliation[c]{Department of Astrophysical Sciences, Princeton University, 4 Ivy Lane, Princeton, NJ, 08544, USA}
\affiliation[d]{School of Natural Sciences, Institute for Advanced Study, 1 Einstein Drive, Princeton, NJ, 08540, USA}

\affiliation[e]{Physics Division, Lawrence Berkeley National Laboratory, 1 Cyclotron Rd, Berkeley, CA, USA}
\affiliation[f]{Center for Cosmology and Particle Physics, Department of Physics, New York University, 726 Broadway, New York, NY 10003, USA}

\affiliation[g]{Department of Physics, Indian Institute of Science Education and Research, Homi Bhabha Road, Pashan, Pune 411008, India}
\usepackage{printlen}

\title{Aemulus $\nu$: Precision halo mass functions in w$\nu$CDM cosmologies}

\abstract{
Precise and accurate predictions of the halo mass function for cluster mass scales in $w\nu{\rm CDM}$ cosmologies are crucial for extracting robust and unbiased cosmological information from upcoming galaxy cluster surveys.
Here, we present a halo mass function emulator for cluster mass scales ({$\gtrsim 10^{13}M_\odot /h$}) up to redshift $z=2$ with comprehensive support for the parameter space of $w\nu{\rm CDM}$ cosmologies allowed by current data. 
Based on the \texttt{Aemulus} $\nu$ suite of simulations, the emulator marks a significant improvement in the precision of halo mass function predictions by incorporating both massive neutrinos and non-standard dark energy equation of state models. 
This allows for accurate modeling of the cosmology dependence in large-scale structure and galaxy cluster studies. 
We show that the emulator, designed using Gaussian Process Regression, has negligible theoretical uncertainties compared to dominant sources of error in future cluster abundance studies. 
Our emulator is \href{https://github.com/DelonShen/aemulusnu_hmf}{publicly available}, providing the community with a crucial tool for upcoming cosmological surveys such as LSST and Euclid.
}

\date{\today}

\begin{document}
\maketitle

\section{Introduction}
The next several years will see a new suite of large cosmological surveys designed to measure the growth of structure and expansion history of the universe and thus probe cosmological parameters, including the evolution of dark energy, the impact of dark matter, and the role of massive neutrinos at significantly higher precision. 
These surveys promise to provide unprecedented data on galaxy clusters, which, as the most massive virialized structures in the universe, provide a unique window into the high-mass, exponentially suppressed tail of the halo mass function  ~\cite{2001ApJ...553..545H, Holder:2001db, Weller:2001gk, Albrecht:2006um, 2011ARA&A..49..409A, 2013PhR...530...87W, Mantz:2014paa, Dodelson:2016wal, SPT:2018njh, DES:2018crd, Zubeldia:2019brr, DES:2020uce, Mazoun:2023kid, DES:2024zpp}. 
For example, current observations of galaxy clusters with optical surveys like the Dark Energy Survey (DES) \cite{DES:2005dhi, DES:2016opl, DES:2016jjg, DES:2020mlx}, Hyper Suprime-Cam \cite{Aihara:2017tri, Sunayama:2023hfm}, and Kilo-Degree Survey (KiDS) \cite{Lesci:2022owx} as well as observations of X-ray emission \cite{2011A&A...534A.109P, Boehringer:2000yh, Burenin:2006td, Pacaud:2015oqr, Ghirardini:2024yni, Bulbul:2024mfj}, and the thermal Sunyaev–Zel’dovich effects \cite{SPT:2014wbo, Planck:2015koh, ACT:2017dgj, SPT:2019hnt, ACT:2020lcv, Raghunathan:2021zfi, SPT:2023via, SPT:2023tib} provide measurements of thousands of galaxy clusters. 
As this number increases dramatically in the era of LSST, Euclid, and the Simons Observatory \cite{LSSTDarkEnergyScience:2012kar, Amendola:2016saw, SimonsObservatory:2018koc, 2013PhR...530...87W}, theoretical models must evolve to match the increasing accuracy of observations. 
In particular, the cosmology dependence of the halo mass function across a wide range of cosmological parameters that account for massive neutrinos and non-standard dark energy models must be precisely calibrated to ensure that theoretical uncertainties remain a negligible source of error and to extract robust and unbiased cosmological information that makes full use of the data.


The halo mass function, which predicts the comoving number density of dark matter halos in a given mass range,  has been a cornerstone of cosmological theory since the spherical collapse model was first used to estimate it in the early work of \Reff{Press:1973iz}. 
This analytic prediction was further formalized in \Reff{Bond:1990iw} with the excursion set approach. 
In these analytic predictions, the halo mass function's dependence on cosmology and redshift was contained primarily within a measure of a halo's rarity: the peak height.
In other words, the mass function was predicted to be ``universal''. 
Over the years, refinements based on increasingly accurate $N$-body simulations have characterized the dependence of the mass function on cosmological parameters \cite{Sheth:1999mn, Sheth:1999su, Jenkins:2000bv,VIRGO:2001szp, Warren:2005ey, Heitmann:2006eu, Diemer:2020rgd, Euclid:2022dbc}, and also have revealed deviations from universality \cite{Reed:2006rw, Lukic:2007fc, Cohn:2007xu, Tinker:2008ff, Courtin:2010gx, 2010MNRAS.403.1353C, 2011ApJ...732..122B}. This non-universality must be explicitly accounted for, especially in the context of upcoming surveys, which will rely on precise modeling of the halo mass function to derive cosmological constraints from cluster abundance data.

Non-universality in the halo mass function can be tackled numerically by building an \textit{emulator}.
Generically, an emulator for a summary statistic is constructed by interpolating between measurements of that summary statistic from a set of sufficiently accurate simulations that carefully sample the desired parameter space.
Once built, the emulator can take input parameters (like cosmology and redshift) and produce predictions for the target statistic with a clear estimate of uncertainty.
Emulators have successfully been constructed for several non-linear summary statistics such as the matter power spectrum \cite{2010ApJ...713.1322L, Heitmann:2013bra, Lawrence:2017ost, Euclid:2020rfv, Moran:2022iwe, DeRose:2023dmk}, 
linear halo bias \cite{Nishimichi:2018etk, McClintock:2019sfj}, 
galaxy clustering and lensing statistics \cite{Wibking:2017slg, Salcedo_2018, Zhai:2018plk, Lange_2021, Kokron:2021xgh, Zennaro:2021bwy, Pellejero-Ibanez:2021tbe, Hadzhiyska:2021xbv, DeRose:2023dmk}, higher-order statistics \cite{StoreyFisher2022,Valogiannis2022, Valogiannis:2023mxf}, and most importantly for this paper, the halo mass function \cite{McClintock:2018uyf, Nishimichi:2018etk, Bocquet:2020tes}.

Building on these foundational studies, here we present a new halo mass function emulator for cluster mass scales $\gtrsim 10^{13} M_\odot / h$ and up to redshift $z=2$, 
with comprehensive support for the full range of  $w\nu{\rm CDM}$ cosmologies allowed by current data. 
Built on the \texttt{Aemulus}~$\nu$ suite of $N$-body simulations outlined in \Reff{DeRose:2023dmk}, our emulator marks a significant improvement in precision by incorporating both massive neutrinos and non-standard dark energy models. 
By using Gaussian Process Regression \cite{10.7551/mitpress/3206.001.0001} to predict halo abundances, we show that the theoretical uncertainties of the emulator are negligible compared to dominant sources of error in upcoming galaxy cluster abundance studies. 
This accuracy of the emulator ensures that its errors remain subdominant for future large-scale cosmological surveys, such as LSST and Euclid.
Following the analytical derivations of the halo mass function in \Refs{Press:1973iz, Bond:1990iw}, we express the fitting function in terms of the linear matter power spectrum. 
This allows the linear matter power spectrum to capture most of the cosmology and redshift dependence of the halo mass function, leaving only the residual non-universal cosmology and redshift dependence to the emulation step. 

The rest of this paper is organized as follows. 
\Sec{sims} reviews the details of the \texttt{Aemulus} $\nu$ suite of simulations and describes how we measure and subsequently fit the halo mass function in each simulation.  \Sec{emu} describes the design and training of our Gaussian process emulator and verifies its accuracy. 
We explore the cosmology dependence of the halo mass function in low and high S8 cosmologies using our emulator in \App{cosmology}.
In \Sec{error}, we construct a model for the theoretical systematic uncertainty introduced by our emulator and demonstrate that this uncertainty due to our emulator will be negligible compared to dominant systematic uncertainties for cluster abundance studies using data from up to LSST Y10. 
We conclude in \Sec{conclusion}.

\section{Measuring the Halo Mass Function from $N$-body Simulation }
\label{sec:sims}
\subsection{The \texttt{Aemulus} $\nu$ Simulation Suite}
\label{sec:simulation}
Cluster cosmology probes halos in the exponentially suppressed high-mass tail of the halo mass function. 
To model this regime precisely, we need N-body simulations with enough resolution and volume to ensure we have a sufficient number of high-mass halos to consider. 
This work uses the \texttt{Aemulus} $\nu$ suite of 150 simulations run in a $w\nu{\rm CDM}$ cosmological parameter space \cite{DeRose:2023dmk}. 
We specify each simulation in this suite with cosmological parameters
\begin{equation}
    \vOm=\{   10^9A_s, n_s, H_0, w_0, \Omega_b h^2, \Omega_c h^2, \Sigma m_\nu\},
    \label{eq:params}
\end{equation}
where the parameters are
\begin{itemize}
    \item $10^9 A_s$: amplitude of scalar fluctuations,
    \item $n_s$: scalar spectral index,
    \item $H_0$: Hubble constant,
    \item $w_0$: dark energy equation of state,
    \item $\Omega_b h^2$: baryon density,
    \item $\Omega_c h^2$: cold dark matter density, and 
    \item $\Sigma m_\nu$: sum of neutrino masses.
\end{itemize}
These simulations are run over a broad parameter space, ensuring that emulators constructed using this suite are accurate over the full range of $w\nu{\rm CDM}$ cosmologies allowed by current data. 
Each simulation has a side length of $L=1050\ h^{-1}{\rm Mpc}$, and jointly evolves the CDM--baryon field (`${\rm cb}$') and the neutrino field, with $N=1400^3$ particles for each species. 
This corresponds to a ${\rm cb}$ particle mass of $3.5\times 10^{10} \left( \frac{\Omega_{\rm cb} } {0.3}  \right)h^{-1}{M_\odot}$. 
This suite of simulations also employs a two-tiered parameter-space design. 
The second tier of 50 simulations, sampled with a Sobol sequence \cite{SOBOL196786}, densely samples the $w\nu{\rm CDM}$ cosmologies currently allowed by the combinations (1) DES Y3 weak lensing and galaxy clustering+BAO+type Ia supernovae and (2) Planck 2018+BAO+type Ia supernovae.
The first tier of 100 simulations, sampled using a Latin hypercube, spans a broader parameter space. 
\neww{More details on the parameter space sampling can be found in \Reff{DeRose:2023dmk}}.
Each simulation is initialized at $z=12$, which is significantly later in cosmic time than previously possible, by using third-order Lagrangian perturbation theory \cite{Michaux:2020yis,Elbers:2020lbn}. 
Doing this mitigates bias in force calculations due to discreteness effects in the particle distribution \cite{Michaux:2020yis}.
We refer the reader to \Reff{DeRose:2023dmk} for more details on this suite of simulations. 

\subsection{Measuring the Mass Function}
In this paper, we define halo mass as 200 times the background CDM + baryon density:
\begin{equation}
\textrm{Halo Mass Definition:}\quad M_{200{\rm cb}}=200\times \textrm{background \underline{CDM + baryon} density}.
\end{equation}
When estimating the halo mass, we count all particles within $R_{\rm 200cb}$ of the halo, even if they are not bound to it.
Namely, these are ``strict spherical overdensity" masses.
We identify halos with the \texttt{ROCKSTAR} halo finder \cite{2013ApJ...762..109B} in 15 redshift snapshots between $z=0$ and $z\approx 2.3$.
For a given snapshot, we bin by mass the identified halos that do not lie within $R_{\rm 200cb}$ of a more massive halo. Namely, we consider only host halos and ignore subhalos.
The mass binning is started at $\log_{10}(200\times M_{\rm particle} )$ rounded up to the nearest tenth. 
For example, if a given box has $\log_{10}(200\times M_{\rm particle}) = 12.85$ we start the mass binning at $\log_{10} M=12.9$. 
We then set the bin width so that we have ten bins per decade: 
\begin{equation}
\textrm{Mass Bin Width}:\quad\Delta \log_{10} M = 0.1.
\label{eq:dlogM}
\end{equation}
After this, we merge the final mass bins associated with the most massive halos until the final mass bin has at least 20 halos\footnote{We neglect redshift snapshots with $<20$ halos that have mass $>200\times M_{\rm particle}$. In our entire suite of simulations, we neglected only a single redshift snapshot of a single tier 1 simulation.}. Then, each mass bin is assigned a ``true mass'' corresponding to the mean mass of halos put into that mass bin. 
Let us define the array of measured number counts in all the redshift snapshots of a specified box as
\begin{equation}
\vN_{\rm true}({\rm Box}):\parbox{4.3in}{\textrm{Array of measured number counts in all redshift snapshots of a specified box.}}\label{eq:Ntrue}
\end{equation}
Once we have measured the mass function for a box at redshift $z$, we can estimate the covariance between mass bins at a fixed redshift for this box using $32^3$ spatial jackknife subvolumes. 
\neww{We demonstrate in \App{jack}} that jackknifing estimates our measurements' sample variance and shot noise \cite{Hu:2002we}. 
We neglect estimating covariance between mass bins in different redshift snapshots. 
Let us define the Jackknifed covariance matrix for measured number counts in all redshift snapshots of a given box as
\begin{equation}
     \mathsf{K}_{\bf N_{\rm true}}({\rm Box}):\parbox{4.3in}{Jacknifed covariance matrix between measured number counts in all redshift snapshot of a specified box.}
     \label{eq:KNtrue}
\end{equation}
\neww{We verify in \Refs{DeRose:2018xdj, DeRose:2023dmk} the convergence of halo abundance measurements from our simulations with respect to resolution, box size, choices in force calculation, and initial conditions.}

\begin{figure}
    \centering
    \includegraphics{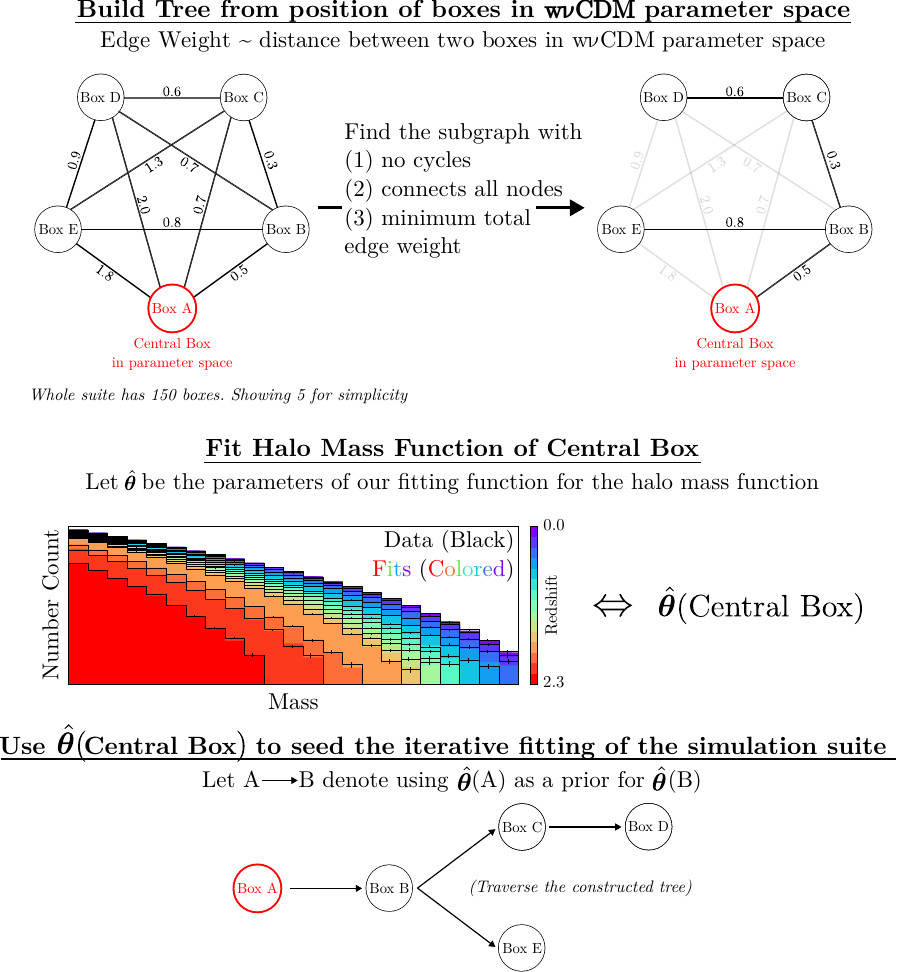}
    \caption{Visualization of our iterative fitting procedure, described in detail in \Sec{fitting}.}
    \label{fig:fitting-visual}
\end{figure}

\subsection{Fitting the Mass Function}
\label{sec:fitting}
We fit the measured abundances $\vN_{\rm true}$ with the functional form of the mass function proposed in Appendix C of \Reff{Tinker:2008ff} modified to depend only on the CDM and baryon mass density field $\rho_{\rm cb}$ as opposed to the whole matter content of CDM, baryons, and massive neutrinos:
\begin{equation}
\frac{dn(M,z)}{dM} = G(\sigma_{\rm cb}(M,z)) \frac{\overline\rho_{\rm cb,0}}M \frac{ d\ln \{\sigma_{\rm cb}(M,z)^{-1}\}}{dM}.
\label{eq:tinker_f}
\end{equation}
This modification to only consider the CDM and baryon content is required because neutrinos do not cluster on halo scales and thus do not contribute to the collapse of dark matter halos \cite{Ichiki:2011ue, Villaescusa-Navarro:2013pva, Costanzi:2013bha}. 
$\overline\rho_{\rm cb,0}$ is the mean CDM and baryon mass density at $z=0$,
\begin{equation}
    \rho_{\rm cb,0} = \frac{3 H_0^2}{8 \pi G}(\Omega_{c} + \Omega_{b}),
\end{equation}
and $\sigma^2_{\rm cb}$ is the variance of the CDM+baryon density field $\rho_{\rm cb}$ smoothed on the Lagrangian radius $R_L = (3 M / 4 \pi \overline{\rho}_{\rm cb,0})^{1/3}$ associated with a mass $M$:
\begin{equation}
\sigma_{\rm cb}^2(R_L[M], z) = \int d\ln k \frac{k^3 P_{\rm cb,lin}(k,z)}{2\pi^2} |\hat W ( k R_L)|^2.
\label{eq:sigM}
\end{equation}
$P_{\rm cb,lin}(k,z)$ is the linear CDM and baryon power spectrum and $\hat W$ is the Fourier transform of the spatial top-hat filter 
\begin{equation}
    \hat W(x) = 3 \frac{\sin(x) - x\cos(x)}{x^3}.
\end{equation}
We compute both $\sigma_{\rm cb}$ and $P_{\rm cb,lin}$ with the Boltzmann code \texttt{CLASS} \cite{2011JCAP...07..034B, 2011JCAP...09..032L}.
The halo multiplicity function $G(\sigma_{\rm cb})$ is defined as
\begin{equation}
G(\sigma_{\rm cb})  = B \left[ \left(\frac{\sigma_{\rm cb}}{e} \right)^{-d} + \sigma_{\rm cb}^{-f} \right] {\rm exp}(-g / \sigma_{\rm cb}^2). \label{eq:G}
\end{equation}
\Refs{Jenkins:2000bv,Warren:2005ey} presented similar fitting functions.
In the halo model, all dark matter lives in a halo \cite{Cooray:2002dia}. Asserting this leads to the normalization 
\begin{equation}
    \int d\ln \sigma^{-1} G(\sigma) = 1.
\end{equation}
This normalization allows us to write $B$ in terms of the remaining free parameters of the halo mass function:
\begin{equation}
    B = 2 \left[e^d g^{-d/2} \Gamma\left(\frac d 2\right) + g^{-f/2} \Gamma\left( \frac f 2 \right) \right]^{-1},
\end{equation}
where $\Gamma(z)$ is the generalized factorial or Gamma function.

We can compute the halo abundance predicted by \Eq{tinker_f}, $\hat{\mathbf N}$, that is suitable to directly compare with the halo abundances we measure from our boxes $\vN_{\rm true}$ from \Eq{Ntrue} by integrating \Eq{tinker_f} in each mass bin and multiplying by the comoving volume of our box $V_{\rm box}$:
\begin{equation}
\hat N = V_{\rm box}\int_{\substack{{\rm mass}\\{\rm bin}}} \frac{dn}{dM}( M) dM.
\label{eq:Nz}
\end{equation}

If we fix the values of $\{d,e,f,g\}$, then \Eq{tinker_f} is a universal mass function.
Our emulator will incorporate non-universality by allowing the parameters $\{d,e,f,g\}$ to vary smoothly as a function of cosmology and redshift.
This allows us to naturally combine the cosmology and redshift dependence of the physically motivated functional form \Eq{tinker_f} while allowing the smooth variation of $\{d,e,f,g\}$ to capture the residual non-universality.

To assert that $\{d,e,f,g\}$ depends smoothly on the cosmological parameters and scale factor, we adopt an iterative fitting approach summarized in \Fig{fitting-visual}.  
\neww{This iterative fitting approach ensures smooth variation of the halo mass function fitting parameters across different cosmologies while naturally handling degeneracies between these parameters. 
By using previously fitted nearby cosmologies as priors for new fits, the method maintains consistency in parameter space while still allowing sufficient flexibility to capture cosmological dependence.}
Let $\mathbf X\sim\mathcal N(\bm \mu, \mathsf K)$ mean that $\mathbf X$ is $n$-dimensional Gaussian random vector with mean $\bm \mu$ and covariance matrix $\mathsf K$:
\begin{equation}
\mathbf X\sim\mathcal N(\bm \mu, \mathsf K)\Leftrightarrow P(\mathbf X) = \sqrt{\frac 1 {(2\pi)^{n}\times  {\rm det}\left[\mathsf K\right]}} {\rm exp} \left\{-\frac 1 2 (\mathbf X - \bm \mu)^T \mathsf K^{-1} (\mathbf X - \bm \mu) \right\}.
\label{eq:grv}
\end{equation}
First, to assert that each $p\in\{d,e,f,g\}$ depends smoothly on scale factor $a$ for a given cosmology $\vOm$ we specify a linear dependence on the scale factor:
\begin{equation}
    p(\vOm,a)=p_0(\vOm) + (a-0.5)p_1(\vOm).
    \label{eq:lindep}
\end{equation}
We found this to be a good approximation for the redshift evolution of $\{d,e,f,g\}$ by individually fitting the parameters in each redshift snapshot and observing a linear dependence.
So, when fitting our suite of simulations, we fit the mass function from all redshift snapshots of a given simulation simultaneously.
Thus, for a given simulation, the parameters we fit are
\begin{equation}
    \vtheta=\{d_0,d_1, e_0, e_1,f_0,f_1,g_0, g_1\}. 
    \label{eq:theta}
\end{equation}
In general, $\vtheta$ are chosen to maximize the posterior
\begin{equation}
    P(\vtheta\mid \hat\vN;{\rm Box}) \propto P(\vtheta;{\rm Box}) \times P(\hat\vN \mid \vtheta; {\rm Box}),\label{eq:post}
\end{equation}
where we build the likelihood using \Eqs{Ntrue}{KNtrue}:
\begin{equation}
(\hat\vN \mid \vtheta; {\rm Box})\sim \mathcal{N}(\vN_{\rm true}({\rm Box}), \mathsf K_{\vN_{\rm true}}({\rm Box})).
\label{eq:likelihood}
\end{equation}
The prior $P(\vtheta;{\rm Box})$ is what we will vary as we go through our iterative fitting.

To start our iterative fitting, we fit parameters of the mass function $\vtheta$ of our central-most box in cosmological parameter space. 
For this central-most box, we first fit the $z=0$ snapshot by setting $d=1.97$\footnote{this specific choice of $d=1.97$ is motivated by Table 4 of \Reff{Tinker:2008ff}} to fix degeneracies and finding $\{e,f,g\}$ that maximize the likelihood \Eq{likelihood}.
We then use these parameters fit from the $z=0$ snapshot to fix one endpoint of the linear dependence \Eq{lindep} of the parameters and fit $\vtheta$ to all redshift snapshots with the prior that
\begin{equation}
\{p(\vOm, a)\mid p \in \{d,e,f,g\}\} \geq 0 .
\end{equation}
By maximizing the posterior \Eq{post} with this prior, we have the fitted parameters $\hat\vtheta({\rm Central\ Box})$.
We then use the fitted $\hat\vtheta({\rm Central\ Box})$ as a prior on the mass function parameters for fitting the next closest few boxes to the central most box 
\begin{equation}
    \vtheta({\rm Nearly\ Central\ Box})\sim \mathcal N(\hat \vtheta({\rm Central\ Box}),  \mathsf K_{\vtheta}).
\end{equation}
We will formally define ``Nearly Central Box" and similar terminology shortly. The parameters for a given nearly central box $\hat\vtheta({\rm Nearly\ Central\ Box})$ are then used as a prior in the same way as $\hat\vtheta({\rm Central\ Box})$. Namely, it is used as a prior on the mass function parameters for fitting the next closest few boxes to this Nearly Central Box:
\begin{equation}
    \vtheta({\rm Near\ Nearly\ Central\ Box})\sim \mathcal N(\hat \vtheta({\rm Nearly\ Central\ Box}),  \mathsf K_{\vtheta})
\end{equation}
This process continues iteratively until we have fit the mass function for all simulations in our suite.
We empirically find that 
\begin{equation}
\mathsf K_{\boldsymbol \theta} = {\rm diag}(10^{-2}, 10^{-2}, 10^{-3}, 10^{-3}, 5\times 10^{-4}, 5\times 10^{-4}, 10^{-4}, 10^{-4})
\end{equation}
is sufficiently relaxed to achieve a sound fit but restrictive enough to wrangle degeneracies between our mass function parameters.

It is left to formalize which boxes are ``nearly central boxes'' and ``near nearly central box'' and so on. To do this, we construct a graph of our boxes to traverse as follows:
\begin{enumerate} 
    \item Standardize each $w\nu{\rm CDM}$ cosmological parameters so that each simulation is labeled with parameters that have comparable characteristic spread (mean zero and unit variance):
    \begin{equation}
        \omega' \equiv \frac{\omega - \langle \omega\rangle}{\sqrt{\langle \omega^2 \rangle - \langle\omega\rangle^2}}\quad \forall \omega\in \{w\nu{\rm CDM}\ \textrm{Cosmological Parameters (\Eq{params})}\}\label{eq:norm}.
    \end{equation}
    \item Build a fully connected graph where each box is a node and edges between two boxes are weighted by the distance in standardized parameter space between those two boxes:
    \begin{equation}
        \texttt{EdgeWeight}({\rm Box\ A}, {\rm Box\ B}) = \sqrt{(\vom'({\rm Box\ A})- \vom'({\rm Box\ B}))^2}
    \end{equation}
    \item Find the subset of this fully connected graph centered on the central-most box in cosmological parameter space that has no cycles, connects all nodes together, and has the minimum total edge weight\footnote{Formally this is known as a minimum spanning tree for which there are well-established algorithms for efficiently finding such a subset \cite{boruuvka1926jistem, kruskal, 6773228, loberman_weinberger, 10.1007/BF01386390}.}.
\end{enumerate}
Thus, we have constructed a graph that formalizes what we mean by ``nearly central boxes'' and so on that we can traverse through to iteratively fit our whole suite of simulations. 

\begin{figure}
    \centering
    \includegraphics{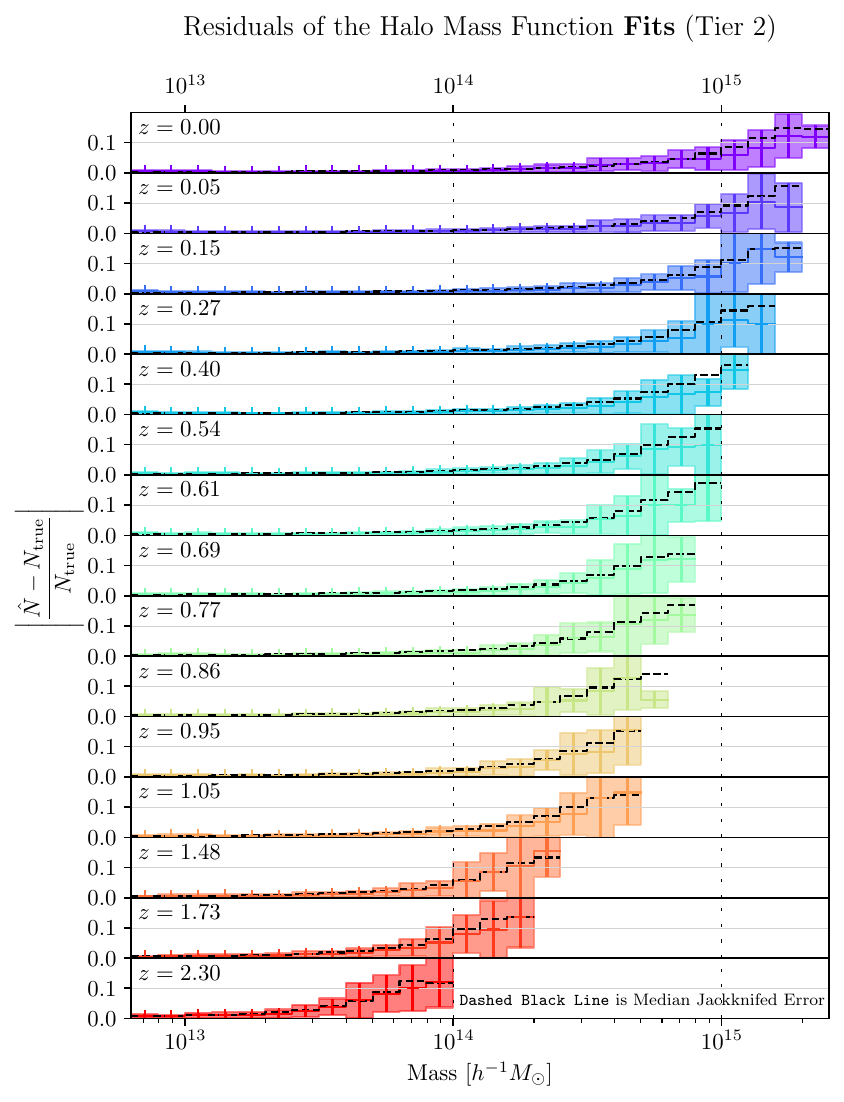}
    \caption{The absolute fractional error on halo abundance from our fits (colored) compared to the median jackknifed error (dashed black) in a given mass bin indicates that for currently allowed $w\nu{\rm CDM}$ cosmologies (tier 2 simulations), the error in abundance from our fits is dominated by the inherent shot noise and sample variance. 
    Bands correspond to $\pm 1\sigma$ from the mean\neww{, where for each mass bin the mean and standard deviation are computed across all tier 2 simulations}.}
    \label{fig:fit-performance}
\end{figure}

\subsection{Fits Performance}
To evaluate the performance of our fitting scheme, we plot in \Fig{fit-performance} the absolute fractional error of the fits versus the actual measured mass function for currently allowed $w\nu{\rm CDM}$ cosmologies (tier 2 simulations)\footnote{We show the equivalent plot on the whole suite of simulations in \App{performance}}. To guide the eye, we plot the median jackknife error in a bin in dashed black. By comparing the fractional errors of our fits with the median jackknife error, we find that the error resulting from the methodology described in \Sec{fitting} is dominated by shot noise and sample variance.

\section{Design of Halo Mass Function Emulator}
\label{sec:emu}
Our emulator uses a Gaussian process (GP), thoroughly treated in \Reff{10.7551/mitpress/3206.001.0001}, to predict the parameters of the fitting function $\vtheta$ (\Eq{theta}) as a function of cosmological parameters (\Eq{params}). 

\subsection{Brief Review of Gaussian Process Regression}
\label{sec:gpreview}
First recall from \Eq{grv} a $n$-dimensional Gaussian random vector 
\begin{equation}
    \vY = [Y_1,Y_2,\dots, Y_n] \sim \mathcal N(\bm\mu, \mathsf K)
\end{equation}
with mean matrix $\bm \mu = [\mu_1, \dots, \mu_n]$ and covariance matrix $\mathsf K$ whose elements we will call $\mathsf K_{ij}$.
For a Gaussian random vector, if we marginalize over any subset of its constituent random variables $\vY^*\subset \vY$, then the remaining $\vY'\equiv\{Y_i \notin \vY^*\}$ will still be described by a Gaussian random vector:
\begin{equation}
    \vY'\sim\mathcal N(\bm \mu', \mathsf K'),\label{eq:marg}
\end{equation}
where $\mathsf K'$ is the relevant submatrix of $\mathsf K$ and $\bm \mu'$ is the relevant subvector of $\bm\mu$.
We have chosen to index the random variables $Y_i$, which make up $\vY$ with integers: $i\in[1,2,\dots,n]$. However, we could have just as easily indexed them with fractions: $i\in [\frac 1n, \frac 2 n,\dots, 1]$. In that case, the mean matrix and covariance matrix are also appropriately relabelled $\bm \mu = [\mu_{1/n}, \mu_{2/n},\dots, \mu_1]$ and $\mathsf K_{i/n, j/n}$. 
We can take the $n\rightarrow \infty$ limit, in which case it would be more appropriate to describe our collection of random variables, $Y_i$, as a scalar function $y(x)$, which is defined for $x\in(0,1]$. 
Correspondingly, the mean and covariance would be more appropriately described as functions $\mu(x)$ and $K(x,x')$. 
This defines a Gaussian process, which we denote as
\begin{equation}
Y \sim \mathcal N(\mu(x), K(x,x'))    
\end{equation}
The property \Eq{marg} generalizes to the statement that any finite number of random variables from our infinite collection of random variables $Y(x)$ are described by a Gaussian random vector.
For example, a vector of random variables like $[Y(x_1), Y(x_2)]$ is still described by a Gaussian random vector:
\begin{equation}
\begin{bmatrix}
    Y(x_1)\\ Y(x_2)
\end{bmatrix}
     \sim\mathcal N\left(\begin{bmatrix}
         \mu(x_1)\\
         \mu(x_2)
     \end{bmatrix}, \begin{bmatrix}
         K(x_1,x_1) & K(x_1,x_2)\\
         K(x_2,x_1) & K(x_2,x_2)
     \end{bmatrix} \right).\label{eq:GPex1}
\end{equation}
This property is the key that enables the use of GPs for regression.
In this example, $x\in(0,1]$ but it is possible to define a GP on an arbitrary interval. 

Now, we schematically describe how GPs are used for regression. Suppose at positions $\vx=[x_1,x_2,\dots, x_n]$ we can measure\footnote{Generally, we would also assume each measurement has some measurement error $\epsilon$. For this simple discussion, we shall neglect this for clarity.} the output of a function we wish to fit, $f(x)$.
One could then assume that $f(\vx)$ is drawn from a Gaussian random vector that is a subset of a Gaussian process:
\begin{equation}
   f(\vx) \sim \mathcal N(0, \mathsf K_\vq)\subset f\sim \mathcal N(0, k_\vq(x,x')),
\end{equation}
where without loss of generality, we have subtracted off the mean so that the GP is centered at zero. 
$\mathsf K$ is determined by the kernel function $k_\vq(x,x')$ with free parameters $\vq$ whose goal is to capture the character of $f(x)$, the function we're fitting:
\begin{equation}
    \mathsf K_{\vq, ij} =  k_{\vq}(x_i, x_j) 
\end{equation}
For example if we had measured $f(x)$ at two point $\vx=[x_1,x_2]$ then we could say 
\begin{equation}
f(\vx)
     \sim\mathcal N\left( \bm 0
     , \begin{bmatrix}
         k_\vq(x_1,x_1) & k_\vq(x_1,x_2)\\
         k_\vq(x_2,x_1) & k_\vq(x_2,x_2)
     \end{bmatrix} \right)\label{eq:GPex2},
\end{equation}
analogous to \Eq{GPex1}. Thus, the parameters of this method of fitting $f(x)$ are the parameters of the kernel function $\vq$. 
These parameters are determined by maximizing the likelihood
\begin{equation}
\argmax_{\vq} \left(-\frac 1 2 f(x_i)(\mathsf K^{-1}_\vq)_{ij} f(x_j) - \frac 1 2 \ln({\rm det}[\mathsf K_\vq])+{\rm const.} \right)\label{eq:gplike}
\end{equation}
where repeated indices are summed over.
In essence, {the success of using GPs for regression boils down to the kernel function's ability to capture the character of the function we are fitting.}  
From here, if we wish to predict $f$ at some points $\vx_*$, then we can draw a sample from the Gaussian random vector 
\begin{equation}
(f(\vx_*)\mid  \vx, f(\vx), \vx_*)\sim (\textrm{GP at some points $\vx_*$} \mid \textrm{GP $=$ the data $f(\vx)$ at points $\vx$})
\end{equation}
where we refer the reader to \Reff{10.7551/mitpress/3206.001.0001} for the explicit form of this distribution.
Though we have focused on GP regression for the simplest 1D case with no measurement noise, the generalization to multiple input and noisy measurements is possible and again discussed in \Reff{10.7551/mitpress/3206.001.0001}. 

\subsection{Specifications of our Gaussian Process Emulator}
The input of our Gaussian process is the cosmological parameters specifying each box \Eq{params}.
To stabilize the training of our Gaussian process, we normalize the inputs so that each input lies between $0$ and $1$:
\begin{equation}
    \omega ' \equiv \frac{\omega - {\rm min}(\omega)}{{\rm max}(\omega) - {\rm min}(\omega)}\quad \forall \omega\in \{w\nu{\rm CDM}\ \textrm{Cosmological Parameters (\Eq{params})}\}.
\end{equation}
Let's first focus on fitting the cosmology and redshift dependence for one of the mass function parameters $t\in \vtheta$.
From \Sec{sims}, we have $n=150$ measurement of this parameter at different cosmologies.
We can use these measurements\footnote{To train the GP, we treat these measurements as ``noiseless" measurements} to model the cosmology dependence of the parameter $t\in\vtheta$ using a Gaussian process as schematically described in \Sec{gpreview}.
As discussed there, the choice of kernel is a critical component in specifying our model.
A popular choice for the kernel is the squared-exponential kernel with parameters $\vq$ where $k_\vq(\vx, \vx')\sim {\rm exp}\left\{-\sum_i(x_i - x_i')^2/\ell_i^2 \right\}$. 
In our emulator, we instead choose to use a more expressive kernel built primarily around the Mat\'ern kernel \cite{10.7551/mitpress/3206.001.0001, matern2013spatial, stein2012interpolation} \neww{which provides additional flexibility to capture the complex relationships between cosmological parameters and mass function behavior.}
\begin{equation}
    k_\vq^{(\vOm)}(\vx, \vx') = A\times\frac{2^{1-\nu}}{\Gamma(\nu)}\left(\frac{|\vx -\vx'|}{C}\times\sqrt{2\nu}\right)^\nu K_\nu\left(\frac{|\vx-\vx'|}{C}\times\sqrt{2\nu}\right) + B \label{eq:GPK}
\end{equation}
where $\Gamma(\nu)$ is the Gamma function and $K_\nu$ is the modified Bessel function of the second kind.
We have chosen the smoothness parameter $\nu=1/2$. The free parameters are the scaling parameters $A$ and $C$ as well as the additive constant to the kernel $B$:
\begin{equation}
\vq = \{A,B,C\}.
\end{equation}
We also model the mean of the Gaussian process as linear in the inputs 
\begin{equation}
\mu_{\{\vw,b\}}(\vOm) = \vw \cdot \vOm + b
\end{equation}
where $\{\vw, b\}$ are the free parameters for this mean model.
Thus, we have specified the architecture of a GP emulator for one mass function parameter as a function of cosmology.

To properly construct a GP emulator that can simultaneously model all the (correlated) mass function parameters $\vtheta$ in \Eq{theta}, we use the framework presented in \Reff{10.5555/2981562.2981582} which extends the covariance in \Eq{GPK} to include covariances between two mass function parameters $t,t' \in \vtheta$, $k_{\vq'}^{(\vtheta)}(t,t')$:
\begin{equation}
    k_{\{\vq, \vq'\}}([\vx, t], [\vx',t']) = k_\vq^{(\vOm)}(\vx,\vx')\otimes k_{\vq'}^{(\vtheta)}(t,t') \label{eq:fullk}
\end{equation}
where $\otimes$ denotes the Kronecker product and $k^{(\vtheta)}_{\vq'}$ is a lookup table with parameters $\vq'$ that contains the halo mass function parameter covariances. 
We model this lookup table as 
\begin{equation}
    k_{\vq'}^{(\vtheta)}(t,t') = (\mathsf D \mathsf D^T + {\rm diag}(\mathbf v))_{t,t'}
\end{equation}
where $\mathsf D$ is a matrix of rank $1$ and $\mathbf v$ is a non-negative vector. 
The free parameters for the covariance between mass function parameters are thus
\begin{equation}
    \vq' = \{\mathsf D, \mathbf v\}.
\end{equation}
We also model the mean for each mass function parameter $t \in \vtheta$ as linear in the inputs 
\begin{equation}
\mu_{\{\vw^{(t)}, b^{(t)}\}}(\vOm) =  \vw^{(t)} \cdot \vOm + b^{(t)}\quad \forall t \in \vtheta\ \textrm{(\Eq{theta})}\label{eq:fullmu}
\end{equation}
where $\{\vw^{(t)}, b^{(t)}\mid t\in \vtheta\}$ are the free parameters for this mean model.
Thus, we have specified the architecture of a GP emulator for all the correlated mass function parameters as a function of cosmology.

We ``train'' the GP specified by \Eqs{fullk}{fullmu}\footnote{To train our model, we use the \texttt{gpytorch} Gaussian Process library \cite{2018arXiv180911165G} but make our final public emulator independent of \texttt{gpytorch} by extracting the trained parameters and using them with our own minimal Gaussian Process library that is constructed to make predictions only.}, e.g., determine the free parameters $\{\vq, \vq', \vw^{(t)}, b^{(t)}\}$ of our GP emulator's mean and kernel components, by maximizing the generalization of likelihood described in \Eq{gplike}.
When training the GP, we use the iterative \texttt{AdamW} optimization algorithm \cite{2017arXiv171105101L} to maximize the likelihood.
The \texttt{AdamW} optimizer is first run with a learning rate of 0.01 for 500 iterations and then decreased to a learning rate of $0.001$ for another 500 iterations.
This training procedure then provides an emulator of the halo mass function parameter as a function of cosmology.

\begin{figure}
    \centering
    \includegraphics{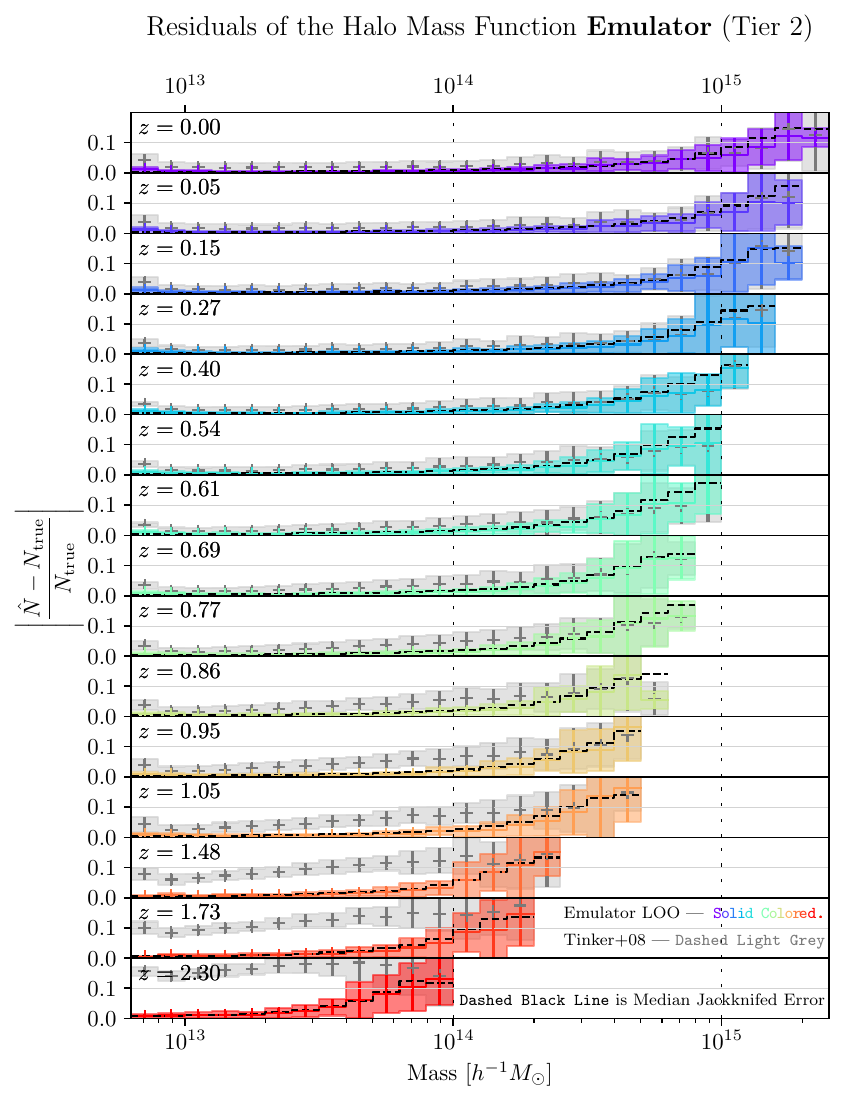}
    \caption{
    The absolute fractional error on halo abundance for a given box from our emulator architecture trained on all but that box (colored) when compared to (1) the median jackknifed error (dashed black) and (2) the error of the Tinker halo mass function modified to depend only on CDM and baryons \cite{Ichiki:2011ue, Villaescusa-Navarro:2013pva, Costanzi:2013bha} (light grey). This indicates that in currently allowed $w\nu{\rm CDM}$ cosmologies, (1) the error in abundance from our emulator is dominated by shot noise and sample variance, and (2) our emulator is significantly more accurate and precise than the Tinker mass function combined with prescriptions that adapt it to $w\nu{\rm CDM}$ cosmologies. 
    We quantitatively model the theoretical systematic error on cluster abundances due to our emulator in \Sec{error} and show in \Fig{ratio_sigemu_sigShapeNoise} that our emulator is sufficiently precise and accurate to be a subleading source of uncertainty in upcoming cluster abundance surveys.     
    Bands correspond to $\pm 1\sigma$ from the mean\neww{, where for each mass bin the mean and standard deviation are computed across all tier 2 simulations}.
    }
    \label{fig:emu-performance}
\end{figure}

\subsection{Emulator Performance}
To evaluate the performance of our emulator, we plot in \Fig{emu-performance} the absolute fractional error between the prediction of an emulator trained on all but one box versus the measured abundances \Eq{Ntrue} of that left out box $|(\hat \vN_{\rm loo}- \vN_{\rm true}) / \vN_{\rm true}|$ for currently allowed $w\nu{\rm CDM}$ cosmologies (tier 2 simulations)\footnote{{We show the equivalent plot on the whole suite of simulations in \App{performance}}}. 
To guide the eye, we also plot (1) the median jackknifed shot noise and sample variance in a given mass bin as a dashed black line and (2) the absolute fractional error when using the Tinker halo mass function presented in \Reff{Tinker:2008ff} modified to depend only on the CDM and baryon content as described \Sec{fitting} and \Refs{Ichiki:2011ue, Villaescusa-Navarro:2013pva, Costanzi:2013bha} in light grey. 
We find that the error of our emulator, similar to the error of the fits shown in \Fig{fit-performance}, is dominated by shot noise and sample variance, meaning additional theoretical uncertainty introduced by our fitting and emulation is negligible.
We quantitatively characterize the theoretical systematic uncertainty on halo abundances due to our emulator in \Sec{error} and show it will be negligible compared to dominant systematic uncertainties for upcoming cluster abundance studies.
Furthermore, by comparing the colored and the light grey bands, we see that in $w\nu{\rm CDM}$ cosmologies currently allowed by data, our halo mass function emulator is significantly more accurate and precise than the standard Tinker mass function \cite{Tinker:2008ff} with prescriptions that adapt it to $w\nu{\rm CDM}$ cosmologies \cite{Ichiki:2011ue, Villaescusa-Navarro:2013pva, Costanzi:2013bha}.
In addition to the decreased precision and accuracy of the Tinker mass function, we also find that the Tinker mass function's response to cosmological parameters differs from our emulator depending on which fiducial cosmology is chosen (\Figs{log_deriv}{log_deriv_lowS8}). 
We explore the implications of this for non-universality more fully in \App{cosmology}.

\section{Emulator Error Quantification and Accuracy Requirement}
\label{sec:error}
In surveys, clusters are binned as a function of some observable, e.g., \texttt{redMaPPer} richness \cite{SDSS:2013jmz}, X-ray luminosity \cite{1986MNRAS.222..323K}, or CMB flux decrement due to the Sunyaev-Zel’dovich effect \cite{1970Ap&SS...7....3S}. 
The mean cluster mass in an observable bin is then determined through some relation between mass and cluster observable. This relation can be calibrated with several techniques, one of the most powerful being weak lensing \cite{2011ApJ...735..118R, 2013PhR...530...87W, SPT:2016gov, Simet:2016mzg, Murata:2017zdo, SPT:2017pbk, DES:2018kma, Murata:2019fxk}.
Generically, this calibration allows us to measure the mean mass of an observable bin $\ln M$ with some uncertainty $\sigma_{\ln M}$.
This uncertainty from the mass-observable calibration leads to an uncertainty in the cluster abundance. 
So, a reasonable accuracy requirement for our halo mass function emulator is that its contribution to uncertainty in cluster abundance is negligible compared to the uncertainty introduced from the mass-observable calibration for upcoming surveys. 
In this section, we first build a model for the theoretical systematic uncertainty on cluster abundances due to our halo mass function emulator. 
We then demonstrate that this theoretical systematic uncertainty due to our emulator is negligible compared to uncertainty from weak lensing mass-observable calibration for up to LSST Y10 data.

\begin{figure}
    \centering
    \includegraphics{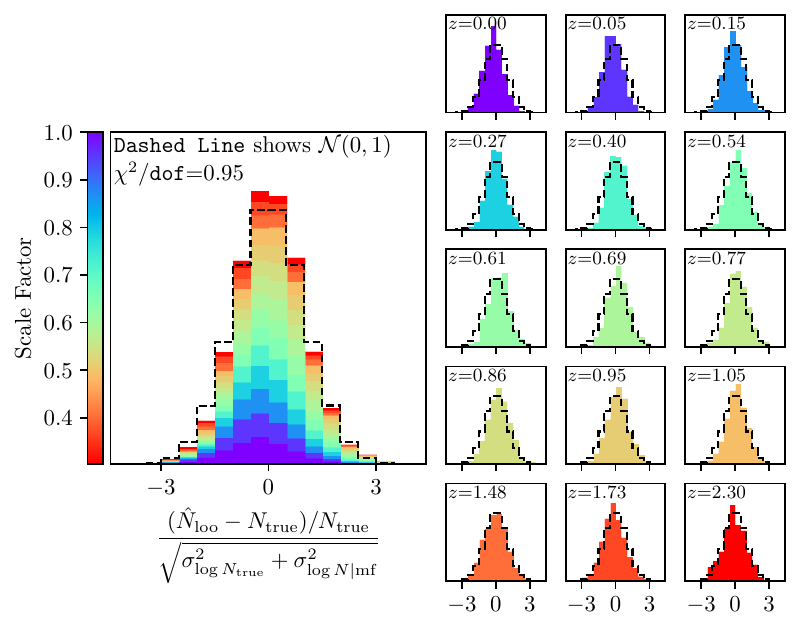}
    \caption{Model of the theoretical systematic uncertainty due to our emulator, $\sigma^2_{\log N\mid \rm mf}$ from \Sec{error} (colored).  The model closely matches that expected for a perfect model of the emulator error (dashed black) when jointly considering residuals in all redshift snapshots in the set of tier 2 simulations (left) but contains minor pathologies when considering a fixed redshift (right).}
    \label{fig:residual_model_distribution}
\end{figure}

\subsection{Theoretical Systematic Uncertainty on Cluster Abundance due to Emulator}
Following \Reff{McClintock:2018uyf}, we model the theoretical systematic uncertainty in abundance measurements due to our emulator $\sigma_{\log N \mid \rm mf}$ as a power law:
\begin{equation}
    \sigma_{\log N\mid \rm mf}(M, a) =A+B\times{\rm exp}\{C\times [\nu(M, a)-3] + D\times [a-0.5] \} ,
    \label{eq:sigmamf}
\end{equation}
where $a$ is the scale factor and $\nu(M,a)$ is the peak height:
\begin{equation}
    \nu(M, a) = \frac{\delta_c}{\sigma_{\rm cb}(M,a)}.
\end{equation}
Here, $\delta_c = 1.686$ is the linear collapse threshold for an Einstein de-Sitter cosmology, and $\sigma_{\rm cb}(M, a)$ is the variance of the smoothed CDM and baryon field described in \Eq{sigM}.
The fitting parameters of $\sigma_{\rm mf}$ are $\{A,B,C,D\}$. 
The fractional residuals in the leave-one-out test shown in \Fig{emu-performance} are due to a sum in quadrature of (1) sample variance and shot noise, estimated by jackknifing, $\mathsf K_{\vN_{\rm true}}$ (\Eq{KNtrue}), and (2) some additional theoretical systematic uncertainty due to the mass function, $\sigma_{\log N \mid \rm mf}$. 
So we can model the total covariance matrix of the fractional residuals $\mathsf K_{\rm frac.res.}$ as
\begin{equation}
    \mathsf K_{\rm frac.res.} = \frac{\mathsf K_{\vN_{\rm true}}}{\vN_{\rm true}\otimes \vN_{\rm true}} +  \mathsf K_{\log N \mid \rm mf}\equiv \mathsf K_{\log N_{\rm true}} + \mathsf K_{\log N\mid \rm mf},
\end{equation}
where 
\begin{equation}
    \mathsf K_{\log N \mid \rm mf} = {\rm diag}(\sigma^2_{\log N\mid \rm mf}(M_1, a), \sigma^2_{\log N \mid \rm mf}(M_2,a),\dots ),
\end{equation}
and $M_i$ is the mean mass of halos in the $i^{\rm th}$ mass bin.
We can fit for $\{A,B,C,D\}$ by modelling the fractional residual in the leave-one-out test shown in \Fig{emu-performance} as 
\begin{equation}
    \left(\frac{\hat\vN_{\rm loo} - \vN_{\rm true}}{\vN_{\rm true}}\, \biggr\lvert\, \{A,B,C,D\}\right)\sim \mathcal N\left(0,  \mathsf K_{\rm frac.res.}\right).
\end{equation}
Maximizing the above likelihood provides a model for the theoretical systematic uncertainty in abundance measurements due to our emulator.

In \Fig{residual_model_distribution}, we plot the distribution of fractional residuals in the leave-one-out test divided by the jackknifed shot noise and sample variance plus our best-fit error model added in quadrature. A perfect error model would yield a normal distribution with unit variance, plotted as a dashed line. We find that when jointly considering residuals in all redshift snapshots in currently allowed $w\nu{\rm CDM}$ cosmologies, our error model yields a distribution that is nearly a Gaussian with unit variance and acceptable reduced $\chi^2$ value of $\chi^2/\texttt{dof}=0.95$.

\begin{figure}
    \centering
    \includegraphics{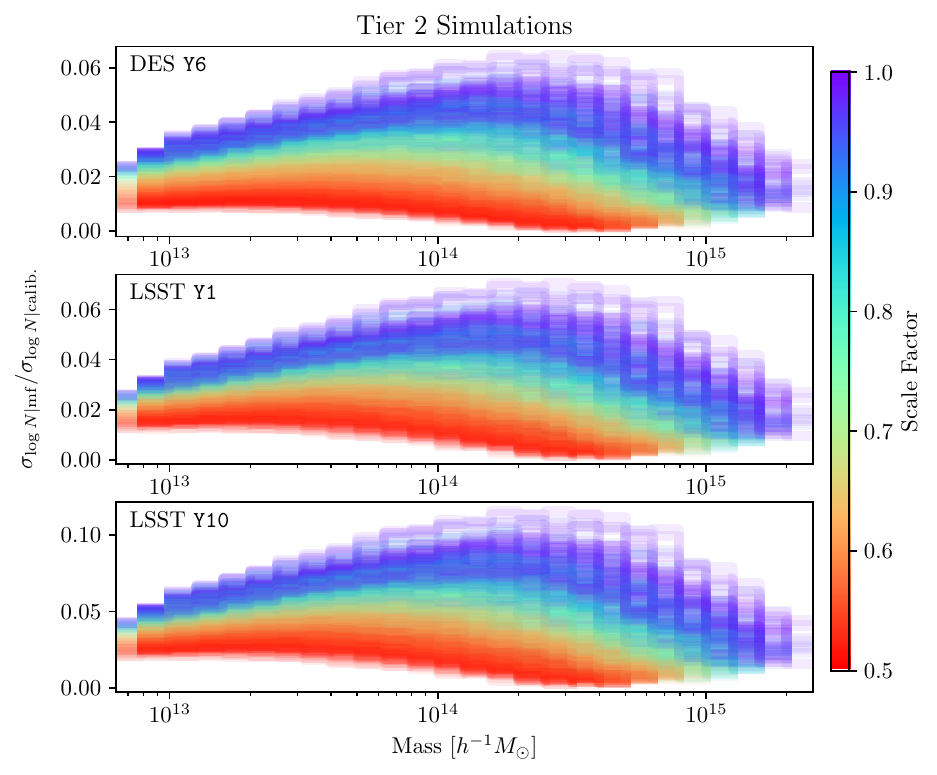}
    \caption{Emulator error compared to the calibration error in the observable--mass relation.
    In cluster abundance studies, the dominant systematic uncertainty in abundance measurements arises from observable--mass relations, which must be calibrated. This calibration introduces some uncertainty $\sigma_{\log N\mid {\rm calib.}}$. Here, we show that the uncertainty on abundances due to our emulator $\sigma_{\log N\mid\rm mf}$ is significantly subleading to uncertainties from weak lensing mass-observable calibration for current and upcoming studies. From top to bottom, the panels show the ratio of these errors in DES Y6 (imminent analysis), LSST Y1 (analysis expected in a few years), and LSST Y10 (note the different y-axis in the final panel). }
    \label{fig:ratio_sigemu_sigShapeNoise}
\end{figure}

\begin{table}
    \centering
    \begin{tabular}{c|ccc}
     & $A\ [h^{-1}M_\odot]^{1.5} $ & $n_0\ [{\rm arcmin}^{-2}]$ & $z_*$ \\
     \hline
        DES Y6 &  $0.7\times 10^{22}$&   $8.4$ & 0.6\\
         LSST Y1 &$0.7\times 10^{22} $  &$10$ & 0.85\\
         LSST Y10 &$0.7\times 10^{22}$  &$27$ & 0.85\\
    \end{tabular}
    \caption{Parameters used in \Sec{shape-noise} for our model of the dominant systematic uncertainty from stacked weak lensing mass calibration, the shape noise, taken from \Refs{McClintock:2018uyf,LSSTDarkEnergyScience:2018jkl}.}
    \label{tab:error_params}
\end{table}

\subsection{Comparison to Systematic Uncertainty in Mass-Observable Calibration}
\label{sec:shape-noise}
We now compute the cluster abundance uncertainty due to weak lensing mass-observable calibration, $\sigma_{\log N\mid{\rm calib.}}$. 
The abundance in a given mass bin can be written as
\begin{equation}
    \log N  = \Delta \log M \rd {\log N}{\log M}
\end{equation}
where as in \Eq{dlogM} we have $\Delta \log_{10} M=0.1$  or 10 mass bins per decade. 
If the bin has a true mass $\log M_b$ but we measure this true mass with some uncertainty $\sigma_{\log M}$, then we can Taylor expand to find the error in abundance due to error in the mass calibration:
\begin{align}
 \nonumber   \log N(\log M_b + \sigma_{\log M}) &= \Delta\log M \rd{\log N}{\log M}\biggr\lvert_{\log M_b + \sigma_{\log M}}\\
\nonumber    &\approx \Delta \log M \left(\rd{\log N}{\log M}+\sigma_{\log M} \rd{^2 \log N}{\log M^2} \right)\biggr\lvert_{\log M_b}\\
    &= \log N_{\rm true} + \sigma_{\log M}\Delta \log M \rd{^2 \log N}{\log M^2}\biggr\lvert_{\log M_b}
\end{align}
From this, we can read off the error in abundance due to an error in the mass-observable calibration, $\sigma_{\log N \mid {\rm calib.}}$, as
\begin{equation}
    \sigma_{\log N \mid {\rm calib.}} = \sigma_{\log M}\times \Delta \log M \rd{^2 \log N}{\log M^2}\biggr\lvert_{\log M_b}
\end{equation}
The dominant source of uncertainty for weak lensing measurements of galaxy cluster masses comes from the intrinsic distribution of galaxy shapes within a catalog \cite{DES:2018kma}. 
This intrinsic variation is called \textit{shape noise}.
So, we approximate the mass--observable calibration uncertainty from weak lensing $\sigma_{\log M}$ as being due only to shape noise\footnote{Note that this is more stringent than \Reff{McClintock:2018uyf}, which approximates the mass--observable calibration uncertainty from weak lensing as being due to shape noise (like is done here) \textit{and} scatter in mass at fixed richness.}:
\begin{equation}
\sigma_{\log M} = \sigma_{\log M}^{\rm SN} =  \frac{\sigma_{M}^{\rm SN}}{M}.
\end{equation}
Following \Reff{McClintock:2018uyf} we model shape noise as
\begin{equation}
    (\sigma_{M}^{\rm SN})^2 = AM^{1/2} (N_c n)^{-1}.
\end{equation}
In the above, $N_c$ is the number of clusters, $A$ is a constant, and $n$ is the lensing source density. As we go to higher redshift, lenses have fewer sources behind them to lens. Thus, our estimate of their masses will become increasingly noisy. This redshift dependence of $\sigma_{\log M}$ comes in as a redshift dependence of the lensing source density $n$, which we model as
\begin{equation}
    n(z) \approx n_0\times {\rm exp}\left\{-\frac 1 2 \frac {z^2}{z_*^2} \right\},
\end{equation}
where $z_*$ is the characteristic redshift where local source density peaks. All together, we have
\begin{equation}
    (\sigma_{M}^{\rm SN})^2 = \frac{ AM^{1/2}}{N_c} \frac{6.3\ {\rm arcmin}^{-2}}{n_0} {\rm exp}\left\{-\frac 1 2 \frac{z^2}{z_*^2} \right\}.
\end{equation}
The numerical values of the parameters of this noise model are taken from \Refs{McClintock:2018uyf, LSSTDarkEnergyScience:2018jkl} and are tabulated in \Tab{error_params}.

In \Fig{ratio_sigemu_sigShapeNoise}, we show that theoretical systematic uncertainty in cluster abundances due to our emulator is significantly smaller than uncertainty due to weak lensing mass-observable calibration for up to LSST Y10 data.
Thus, theoretical systematic uncertainty due to our halo mass function emulator will be a negligible source of error for upcoming cluster abundance studies.

\section{Conclusion}
\label{sec:conclusion}
Precision predictions of the halo mass function in $w\nu{\rm CDM}$ cosmologies are crucial for extracting robust cosmological information from upcoming galaxy cluster surveys.
In this paper, we have presented a halo mass function emulator suitable for cluster mass scales $M_{200cb}\gtrsim 10^{13}\ M_\odot / h$ and redshifts $z\leq 2$ for a broad range of $w\nu$CDM cosmologies. 
The emulator is constructed with the \texttt{Aemulus} $\nu$ suite of $N$-body simulations which span the broadest $w\nu{\rm CDM}$ parameter space ever used in a single suite of simulations. 
We use a fitting function for the halo mass function presented in \Reff{Tinker:2008ff} modified to depend on only the CDM and baryon components \cite{Ichiki:2011ue, Villaescusa-Navarro:2013pva, Costanzi:2013bha} to fit the measured halo abundances in each simulation in this suite. 
A Gaussian process is then trained on the measured parameters of this fitting function to predict the cosmology dependence of these parameters.
This design allows us to naturally combine the inherent cosmology and redshift dependence of universal halo mass functions derived from physically motivated models while directly capturing residual non-universality through the Gaussian process. 
We find (\Fig{emu-performance}) that for currently allowed $w\nu{\rm CDM}$ cosmologies, our emulator is significantly more accurate and precise than the Tinker mass function \cite{Tinker:2008ff} combined with prescriptions that adapt it to $w\nu{\rm CDM}$ cosmologies  \cite{Ichiki:2011ue, Villaescusa-Navarro:2013pva, Costanzi:2013bha}.
We quantify the theoretical systematic uncertainty on cluster abundances due to our halo mass function emulator. 
This is compared with uncertainties due to weak lensing mass--observable calibration from DES Y6, LSST Y1, and LSST Y10 data. 
We find (\Fig{ratio_sigemu_sigShapeNoise}) that the theoretical systematic uncertainty on cluster abundances due to our emulator will be negligible compared to uncertainty due to mass-observable calibration for upcoming cluster abundance studies. 
We make our emulator for the halo mass function \href{https://github.com/DelonShen/aemulusnu_hmf}{publicly available} and encourage its use in future cosmological analyses.

\section*{Acknowledgments}

We thank Federico Bianchini, Adam Mantz, Emmanuel Schaan, and Chun-Hao To for helpful discussions.
This work received support from the U.S. Department of Energy under contract number DE-AC02-76SF00515 to SLAC National Accelerator Laboratory. 
D.S. is additionally supported by the National Science Foundation Graduate Research Fellowship under Grant No. DGE-2146755. N.K. acknowledges support from NSF award AST-2108126 and from the Fund for Natural Sciences of the Institute for Advanced Study. J.D.~is supported by the Lawrence Berkeley National Laboratory Chamberlain Fellowship. JT is supported by National Science Foundation grant 2009291.
This research used resources of the National Energy Research Scientific Computing Center, a DOE Office of Science User Facility supported by the Office of Science of the U.S. Department of Energy under Contract No. DE-AC02-05CH11231. Some of the computing for this project was performed on the Sherlock cluster. We thank Stanford University and the Stanford Research Computing Center for providing computational resources and support that contributed to these research results.  This work used \texttt{Stampede2} at the Texas Advanced Computing Center and \texttt{Bridges2} at the Pittsburgh Supercomputing Center through allocation PHY200083 from the Extreme Science and Engineering Discovery Environment (XSEDE) \cite{towns2014xsede}, which was supported by National Science Foundation grant number 1548562.

\appendix

\section{Cosmology Dependence of the Halo Mass Function in Low and High S8 Cosmologies}
\label{app:cosmology}

\begin{figure}[p]
    \centering
    \includegraphics{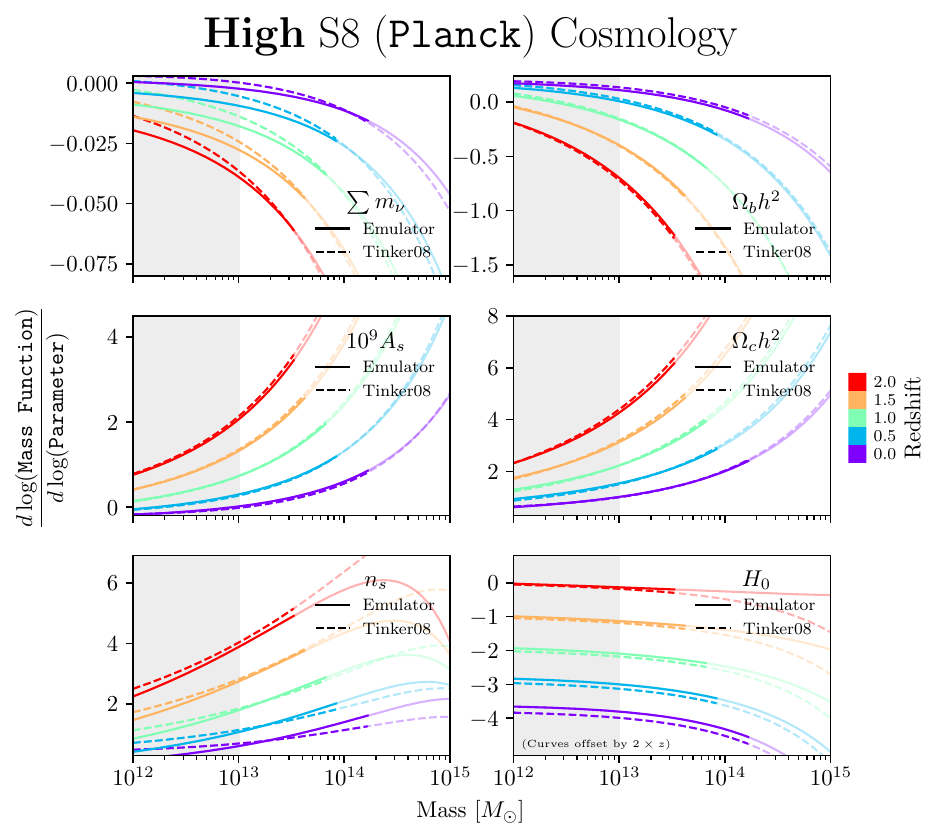}
    \caption{For a high S8 (Planck \cite{Planck:2018vyg}) cosmology, the logarithmic derivative of the mass function with respect to cosmological parameters, and specifically $n_s$, significantly changes between our non-universal halo mass function emulator (solid) and the standard universal Tinker mass function \cite{Tinker:2008ff} modified to depend on only the CDM and baryon content \cite{Ichiki:2011ue, Villaescusa-Navarro:2013pva, Costanzi:2013bha} (dashed) in the mass range where we have a significant number of halos to consider within our simulation (darker). 
    This can be compared to a low S8 (DES Y3) cosmology shown in \Fig{log_deriv_lowS8}, where the logarithmic derivatives differ less between our halo mass function emulator and the Tinker mass function. 
    The difference in cosmology dependence of the halo mass function in a low or high S8 cosmology is because more clusters exist in a high S8 cosmology.
    Therefore, the halo mass function on cluster mass scales has a greater dynamic range, allowing for greater cosmology dependence and impact from non-universality.
    Lighter parts of the lines correspond to where the jackknifed error on cluster abundance measurements from our simulations exceeds 1\%, and the light grey region corresponds to the region that is $\lesssim 200\times M_{\rm particle}$.}
    \label{fig:log_deriv}
\end{figure}

\begin{figure}[p]
    \centering
    \includegraphics{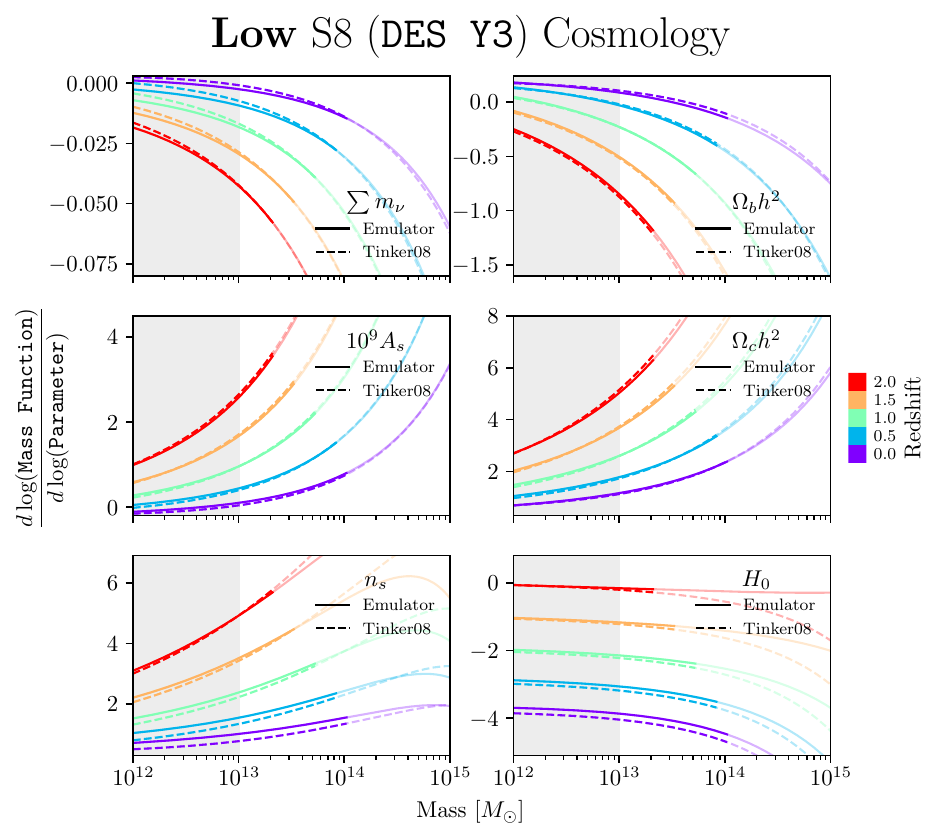}
    \caption{
    For a low S8 (DES Y3 \cite{DES:2021wwk}) cosmology, the logarithmic derivative of the mass function with respect to cosmological parameters does not change significantly between our non-universal halo mass function emulator (solid) and the standard universal Tinker mass function \cite{Tinker:2008ff} modified to depend on only the CDM and baryon content \cite{Ichiki:2011ue, Villaescusa-Navarro:2013pva, Costanzi:2013bha} (dashed) in the mass range where we have a significant number of halos to consider within our simulation (darker). 
    This can be compared to a high S8 (Planck) cosmology shown in \Fig{log_deriv}, where the logarithmic derivatives differ much more between our halo mass function emulator and the Tinker mass function. 
    The difference in cosmology dependence of the halo mass function in a low or high S8 cosmology is because more clusters exist in a high S8 cosmology.
    Therefore, the halo mass function on cluster mass scales has a greater dynamic range, allowing for greater cosmology dependence and impact from non-universality.
    Lighter parts of the lines correspond to where the jackknifed error on cluster abundance measurements from our simulations exceeds 1\%, and the light grey region corresponds to the region that is $\lesssim 200\times M_{\rm particle}$.}
    \label{fig:log_deriv_lowS8}
\end{figure}

In a high S8 (Planck \cite{Planck:2018vyg}) cosmology, the cosmology dependence of the halo mass function on cluster mass scales, shown in \Fig{log_deriv}, changes significantly between the universal Tinker halo mass function modified to depend only on the CDM and baryon content as described \Sec{fitting} and \Refs{Ichiki:2011ue, Villaescusa-Navarro:2013pva, Costanzi:2013bha} and our non-universal halo mass function emulator. 
Most notable is the change in the character of $n_s$ dependence\footnote{The fact that changes in the spectral index $n_s$ capture a significant amount of non-universality is not a new idea in the literature. For example, \Reff{Ondaro-Mallea:2021yfv} found that an analogous quantity, the local power spectrum slope defined as the slope of $\sigma(R)$ at the Lagrangian radius of a halo of mass $M$, was a primary driver of non-universality. However, later on, \Reff{Guo:2024had} found that neither the local slope of the power spectrum nor $n_s$ played any significant role in non-universality for the halo mass function at $z=0$ and instead utilized deep learning to identify three other variables that are necessary and sufficient to predict the HMF at $z=0$ to sub percent accuracy.}. 
This differs from a low S8 (DES Y3 \cite{DES:2021wwk}) cosmology where the cosmology dependence of the halo mass function on cluster mass scales, shown in \Fig{log_deriv_lowS8}, changes less between the universal Tinker halo mass function and our non-universal halo mass function emulator in comparison to a high S8 cosmology. 
This difference between low and high S8 cosmologies occurs because there are more clusters in a high S8 cosmology.
Thus, the halo mass function has a greater dynamic range for cluster mass scales, which allows for increased cosmology dependence and impact from non-universality.


\section{\neww{Validating Jackknife Estimates of Halo Abundance Measurement Covariance}}
\label{app:jack}
\begin{figure}
    \centering
    \includegraphics[width=\linewidth]{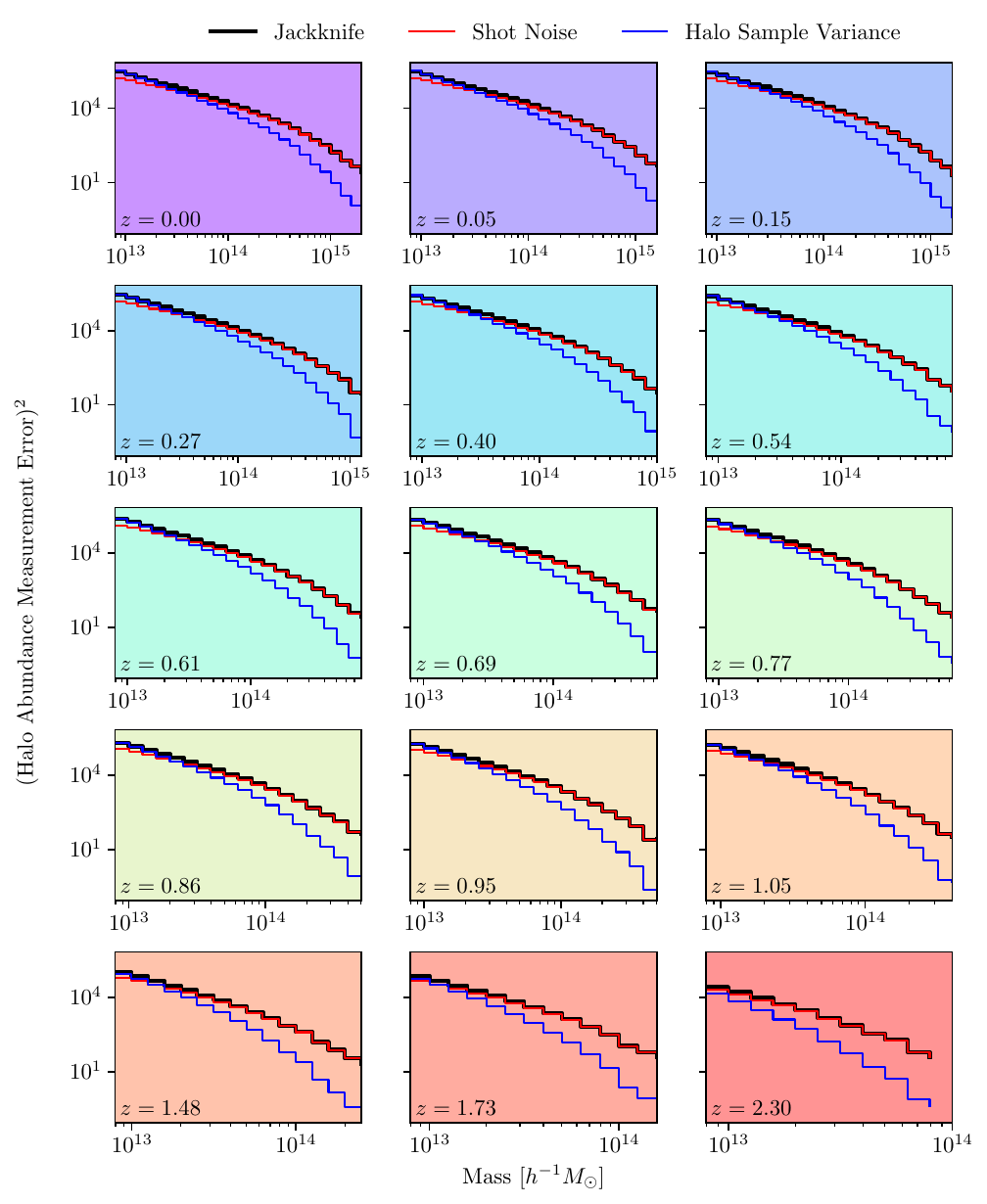}
    \caption{Our jackknife estimated errors of the halo abundance measurements from our simulations (black, \Eq{KNtrue}) are reasonable when compared with halo sample variance (blue, \Eq{hsv}), which should dominate at lower masses, and shot noise (red, \Eq{shot}), which should dominate at higher masses.
    }
    \label{fig:jackknife_verify}
\end{figure}

In this appendix, we verify that jackknifing yields reasonable estimates of the halo abundance measurement covariance (\Eq{KNtrue}) by comparing the jackknife estimated variance with analytical estimates of the halo sample variance and shot noise\cite{Hu:2002we}. 
Assume the number density fluctuation $\delta n(M)$ of high mass halos with mass $M$ in some range $dM$ is related to the underlying linear mass field $\delta$ by a linear bias parameter
\begin{equation}
    n(M) = \bar n(M) + \delta n(M)= \bar n(M)(1 + b(M) \delta).
\end{equation}
So the average number density within our box is 
\begin{equation}
    \langle n(M)\rangle _{\rm box} = \frac 1 {L_{\rm box}^3}\int_{\rm box} d^3x\ (\bar n + \delta n).
\end{equation}
The sample variance of this measurement then will be
\begin{align}
    \sigma^2_{\rm SV} = \langle n(M)^2\rangle_{\rm box} - \langle n(M)\rangle_{\rm box}^2 &= b^2 \bar n(M)^2 \int\frac{d^3 k}{(2\pi)^3} P_{\rm cb,lin}(k, z) \prod_{i=x,y,z} \sinc^2\left(\frac{ k_i L_{\rm box}}{2} \right).\label{eq:hsv}
\end{align}
We approximate $b(M)$ in our estimate of the halo sample variance with the halo bias introduced in \Reff{tinker10}.
On top of the sample variance is the shot-noise variance.
\begin{equation}
    \sigma^2_{\rm Shot} = \bar n(M).\label{eq:shot}
\end{equation}
The sample variance $\sigma^2_{\rm SV}$ dominates for low masses, while the shot noise $\sigma_{\rm Shot}^2$ dominates for high masses \cite{Hu:2002we}.
In \Fig{jackknife_verify} we verify that our jackknife estimated errors on halo abundance are reasonable when compared with analytical estimates of halo sample variance and shot noise for the central most box.

\section{Complete Performance Plots}
\label{app:performance}
\begin{figure}
    \centering
    \includegraphics{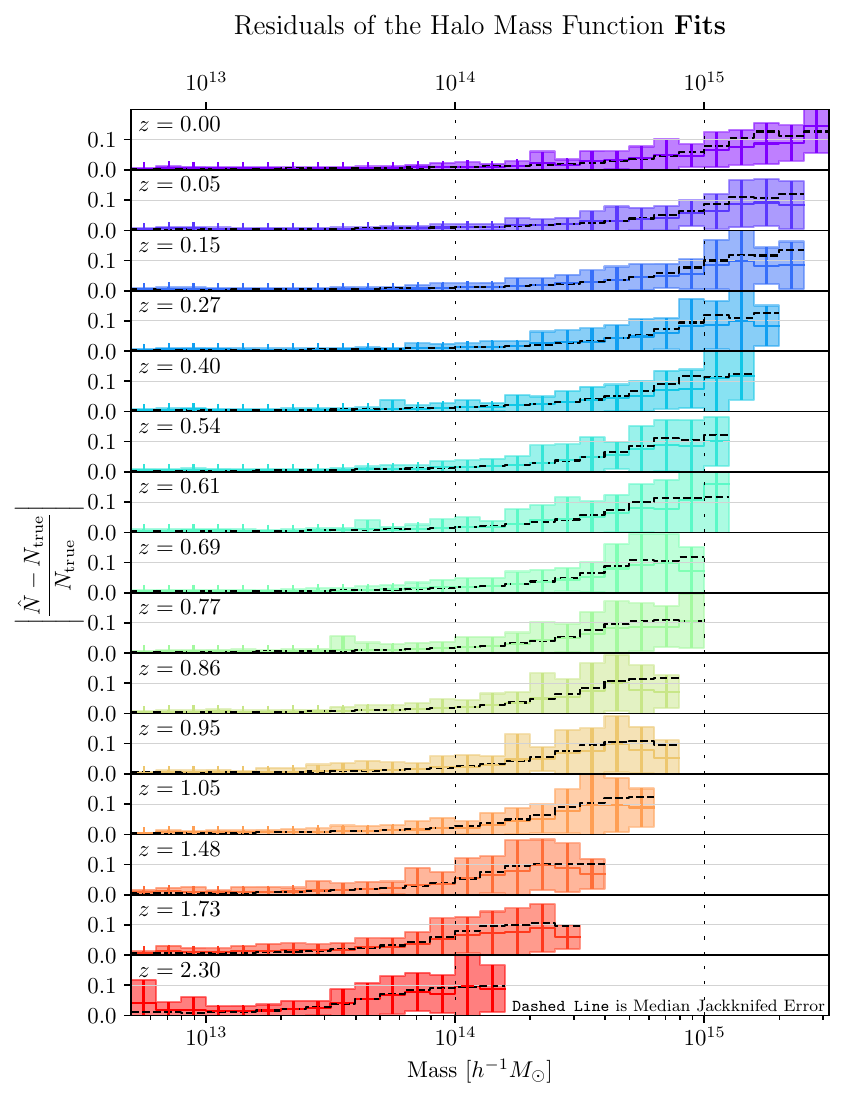}
    \caption{The absolute fractional error on halo abundance from our fits (colored) compared to the median jackknifed error (dashed black) in a given mass bin. 
    This plot is analogous to \Fig{fit-performance} but for the whole suite of simulations instead of the tier 2 simulations that are of primary interest.
    Bands correspond to $\pm 1\sigma$ from the mean\neww{, where for each mass bin the mean and standard deviation are computed across all simulations}.
    }
    \label{fig:fit-performance-all}
\end{figure}

\begin{figure}
    \centering
    \includegraphics{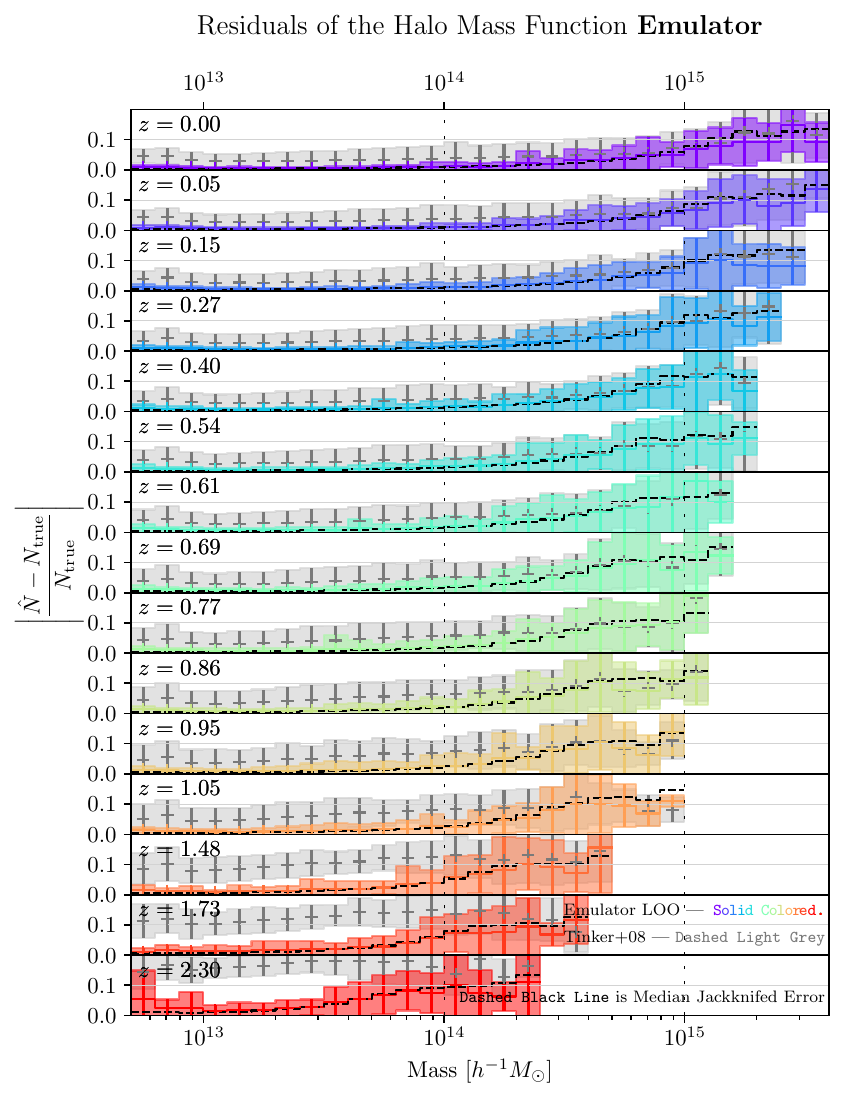}
    \caption{
    The absolute fractional error on halo abundance for a given box from our emulator architecture emulator trained on all but that box (colored) compared to the median jackknifed error (dashed black) and error of the Tinker halo mass function modified to depend only on CDM and baryons \cite{Ichiki:2011ue, Villaescusa-Navarro:2013pva, Costanzi:2013bha} (light grey) in a given mass bin. This plot is analogous to \Fig{emu-performance} but for the whole suite of simulations instead of the tier 2 simulations that are of primary interest.
    Bands correspond to $\pm 1\sigma$ from the mean\neww{, where for each mass bin the mean and standard deviation are computed across all simulations}.}
    \label{fig:emu-performance-all}
\end{figure}

In this appendix, we show the performance plots of our fits and emulator for the whole suite of simulations instead of only the tier 2 simulations. 
As described in \Sec{simulation}, the \texttt{Aemulus} $\nu$ suite of simulations employs a two-tiered parameter space design. 
The primary tier of interest is tier 2, which spans the range of $w\nu{\rm CDM}$ cosmologies allowed by current data. 
Namely, (1) DES Y3 weak lensing and galaxy clustering+BAO+type Ia supernovae and (2) Planck 2018+BAO+type Ia supernovae. 
So, these are the cosmologies where the accuracy and precision of our emulator are of most interest. 
Because of this, our performance plots throughout this paper have been restricted to only performance on the tier 2 simulations. 
In \Fig{fit-performance-all}, we plot the absolute fractional error of the fits versus the actual measured mass function for all the simulations. 
This is the generalization of \Fig{fit-performance}. 
In \Fig{emu-performance-all}, we plot the absolute fractional error between the prediction of an emulator trained on all but one box versus the measured abundances \Eq{Ntrue} of that left-out box for all the simulations.
This is the generalization of \Fig{emu-performance}.

\bibliography{main}
\bibliographystyle{jhep}

\end{document}